\documentclass[journal]{IEEEtran}
\usepackage{bbm}
\usepackage{amsfonts}
\usepackage{mathrsfs}
\usepackage {amssymb}
\usepackage {amsmath}
\usepackage{amsthm}
\usepackage{latexsym}
\usepackage{makecell,rotating}
\usepackage[style=plaintop]{floatrow}
\usepackage{booktabs}
\ifCLASSINFOpdf
\else
\fi
\begin{document}
%
\title{A Class of Two-Weight and Three-Weight Linear Codes and Their Duals}
%
%
%

\author{Li Liu, Xianhong Xie and Lanqiang Li}
\maketitle

\begin{abstract}
The objective of this paper is to construct a class of linear codes with two nonzero weights and three nonzero weights by using the general trace functions, which weight distributions has been determined. These linear codes contain some optimal codes, which meets certain bound on linear codes. The dual codes are also studied and proved to be optimal or almost optimal. These codes may have applications in authentication codes, secret sharing schemes and strongly regular graphs.
\end{abstract}


%
\IEEEpeerreviewmaketitle

\section{Introduction}
%
%
%
%
\IEEEPARstart{T}{hrough} this paper, let $p$ be prime and $q=p^{s}$, where $s$ is a positive integer. An $[n,k,d]$ code $C$ over $F_{p}$ is a k-dimension subspace of $F^{n}_{p}$ with minimum Hamming distance $d$. Let $A_{i}$ denote the number of codewords with Hamming weight $i$ in $C$, then $(1,A_{1},\cdots,A_{n})$ is called the weight distribution of $C$.

Griesmer Bound is a generalization of the Singleton Bound and is different from other upper bounds. This one only applies to linear codes. Therefore, Griesmer Bound is presented in the following lemma[11].

\noindent\textbf{Lemma 1.1.} Let $C$ be an $[n,k,d]$ code over $F_{q}$, with $k\geq 1$. Then
$$n\geq \sum^{k-1}_{i=0}\lceil \frac{d}{q^{i}}\rceil.$$

The Griesmer Bound gives a lower bound on the length of a code over $F_{q}$ with a specified dimension $k$ and minimum distance $d$. An $[n,k,d]$ code $C$ is called optimal if parameters $[n,k,d]$ meet this bound. An $[n,k,d]$ code $C$ is called almost optimal if $[n,k,d+1]$ meet this bound.

Let $D=\{d_{1},d_{2},\cdots, d_{n}\} \subseteq F_{q}^{*}$, then a linear code over $F_{p}$ of length $n$ is
\[C_{D}=\{(Tr^{s}_{1}(\beta d_{1}),Tr^{s}_{1}(\beta d_{2}),\cdots,Tr^{s}_{1}(\beta d_{n}))\},\]
where $Tr^{s}_{1}$ is the trace function from $F_{q}$ onto $F_{p}$. $D$ is called defining set of this code $C_{D}$.
The selection of $D$ directly affects the constructed linear codes. So we can obtain linear codes with few weights by the proper selection of $D$ [5],[12],[13],[15],[17].
In addition, from the another view of $D$, let $D=\{x\in F_{q}:Tr^{s}_{1}(f(x))=0\}$ and $f(x)\in F_{q}(x)$. The previous work focused on changing the function $f(x)$ or generalizing the defining set of $D$, i.e, $f(x)=x^{2^{h}+1}$[3], $f(x)=x^{\frac{q-1}{7}}(3|n)$[9]. Ding and Ding[4] gave the weight distributions of $C_{D}$ for the case $f(x)=x^{2}$ and proposed an open problem on calculating the weight distributions of $C_{D}$ for general planar functions. Zhou et al.[7] solved this problem by the quadratic bent functions. Tang et al.[10] also settled the problem by the weakly quadratic bent functions. A class of two-weight and three-weight linear codes with the general trace functions has given by Tang et al.[6]. Carlet and Ding[1] presented the minimum distance of $C$ and $C^{\bot}$ for the case $f(x)=\alpha x^{p^{k}+1}+\beta x$. For the case of $\frac{s}{gcd(s,k)}$ being odd, Yuan et al.[2] gave the weight distributions. However, the problem of constructing weight distributions for $\frac{s}{gcd(s,k)}$ being even is underdeveloped.

In this paper, motivated by the research work in [1] and [3], we use the more general method to construct linear codes with two and three weights. New parameters and weight distributions of such codes are determined. Some of the linear codes in this paper are optimal. Besides, linear codes with three-weight and two-weight of this paper may have applications in secret sharing schemes [16] and authentication codes [15].


\section{Linear codes with two weights and three weights}
In this section, we only describe the linear codes and introduce their parameters by two theorems. The proofs of their parameters will be presented later.

Let $s=2m$, $m=et$ and $q=p^{2m}$, where $m$, $t$ and $e$ are positive integers. Define

\begin{equation*}
  D_{1}=\{x\in F^{*}_{q}: Tr^{s}_{e}(x^{p^{m}+1})=0\},
\end{equation*}

where $Tr^{s}_{e}(x)=\sum\limits_{i=0}^{\frac{s}{e}-1}x^{p^{ei}}$ is the general trace function. Let $D_{1}=\{d_{1},d_{2},\cdots,d_{n_{1}}\}$ and $n_{1}=|D_{1}|$, we have the linear code
\begin{equation}\label{1}
  C_{D_{1}}=\{c_{\beta}: \beta \in F_{q}\}
\end{equation}
where $c_{\beta}=(Tr^{s}_{1}(\beta d_{1}),Tr^{s}_{1}(\beta d_{2}),\cdots,Tr^{s}_{1}(\beta d_{n_{1}}))$.

\noindent\textbf{Theorem 2.1.} Let $s=2m$, $e|m$ and $e<m$. The code $C_{D_{1}}$ defined in Equation (1) is a two-weight linear code with parameters $[p^{2m-e}+p^{m-e}-p^{m}-1, 2m, (p^{2m-e-1}-p^{m-1})(p-1)]$, whose weight distribution is listed in \noindent\textbf{Table I}.
\begin{table}
\caption{THE WEIGHT DISTRIBUTION OF THE CODE OF THEOREM 2.1 }
\begin{tabular}{@{}llr@{}}\toprule
Weight $w$     &  Multiplicity $A_{w}$ \\ \midrule
$p^{2m-e}-p^{2m-e-1}$  &  $p^{2m-e}-(p^{e}-1)p^{m-e}-1$ \\
$(p^{2m-e-1}-p^{m-1})(p-1)$  &  $(p^{e}-1)(q+p^{m})/p^{e}$ \\ \bottomrule
\end{tabular}
\end{table}

\noindent\textbf{Example 2.1.} Let $(m,e)=(2,1)$ and $p=3$. Then the code $C_{D_{1}}$ has parameters $[20,4,12]$ and weight enumerator $1+20x^{18}+60x^{12}.$ This code is optimal due to the Griesmer bound since the optimal linear code over $F_{3}$ with length 20 and dimension 4 has minimum weight 12.

\noindent\textbf{Example 2.2.} Let $(m,e)=(2,1)$ and $p=5$. Then the code $C_{D_{1}}$ has parameters $[104,4,80]$ and is almost optimal, while the optimal linear code has parameters $[104,4,81].$

It is observed that the weights in the code $C_{D_{1}}$ have a common divisor $p-1$. This indicates that the code $C_{D_{1}}$ may be punctured into a shorter one whose weight distribution can be derived from that of the original code $C_{D_{1}}$. This will be done as follows.

Note that for any $a\in F^{*}_{p}$, $Tr^{s}_{e}((ax)^{p^{m}+1})=a^{2}Tr^{s}_{e}(x^{p^{m}+1})$. We can select a subset $\overline{D_{1}}$ of $D_{1}$ such that
\begin{equation}\label{2}
  D_{1}=(F_{p}^{*})\overline{D_{1}}=\{ab:~a\in F_{p}^{*}~,~b\in \overline{D_{1}}\},
\end{equation}
 where $\frac{b_{i}}{b_{j}}\notin F_{p}^{*}$ for every pair of distinct elements  $(b_{i}, b_{j}) \in\overline{D_{1}}^{2}$. Hence, the parameters and weight distributions of the $C_{\overline{D_{1}}}$ are given in the following corollary.

\begin{table}
\caption{THE WEIGHT DISTRIBUTION OF THE CODE OF COROLLARY 2.2 }
\begin{tabular}{@{}llr@{}}\toprule
Weight $w$     &  Multiplicity $A_{w}$ \\ \midrule
$p^{2m-e-1}$  &  $p^{2m-e}-(p^{e}-1)p^{m-e}-1$ \\
$p^{2m-e-1}-p^{m-1}$  &  $(p^{e}-1)(q+p^{m})/p^{e}$ \\ \bottomrule
\end{tabular}
\end{table}
\noindent\textbf{Corollary 2.2.} Let $s=2m$, $e|m$ and $e<m$. Let $\overline{D_{1}}$ be defined in (2). Then the code $C_{\overline{D_{1}}}$ is a two-weight linear code with parameters $[\frac{p^{2m-e}+p^{m-e}-p^{m}-1}{p-1},2m,p^{2m-e-1}-p^{m-1}]$ whose weight distribution is listed in \noindent\textbf{Table II}.

\noindent\textbf{Example 2.3.} Let $(m,e)=(3,1)$ and $p=3$. the code $C_{D_{1}}$ has parameters $[224,6,144].$ Note that the code constructed is not optimal, since an optimal [224, 6] code has minimum weight 147. The code $C_{\overline{D_{1}}}$ has parameters $[112,6,72]$. This code is optimal due to the Griesmer bound since the optimal linear code with length 112 and dimension 6 has minimum weight 72.

\noindent\textbf{Example 2.4.} Let $(m,e)=(2,1)$ and $p=5$. Then the code $C_{D_{1}}$ has parameters $[104,4,80]$ and is almost optimal. But the code $C_{\overline {D_{1}}}$ has parameters $[26,4,20]$. This code is optimal.

Define $D_{2}=F^{*}_{q}$, let $D_{2}=\{d_{1},d_{2},\cdots,d_{n_{2}}\}$, where $n_{2}=p^{2m}-1$. We define a linear code of length $n_{2}$ over $F_{p}$ by
\begin{equation}\label{3}
  C_{D_{2}}=\{c_{(\beta,\gamma)}:\beta\in F_{q},\gamma\in F_{p^{m}}\},
\end{equation}
where
$$c_{(\beta,\gamma)}=((Tr^{s}_{1}(\beta d_{1})+Tr^{m}_{1}(\gamma d_{1}^{p^{m}+1})),(Tr^{s}_{1}(\beta d_{2})+$$
$$Tr^{m}_{1}(\gamma d_{2}^{p^{m}+1})),\cdots,(Tr^{s}_{1}(\beta d_{n_{2}})+Tr^{m}_{1}(\gamma d_{n_{2}}^{p^{m}+1}))).$$

\noindent\textbf{Theorem 2.3.} Let $s=2m$. Then the code $C_{D_{2}}$ defined in (3) is a three-weight linear code with parameters $[p^{2m}-1,3m]$ whose weight distribution is listed in \noindent\textbf{Table III}.

\begin{table}
\caption{THE WEIGHT DISTRIBUTION OF THE CODE OF THEOREM 2.3 }
\begin{tabular}{@{}llr@{}}\toprule
Weight $w$     &  Multiplicity $A_{w}$ \\ \midrule
$p^{2m-1}(p-1)$  &  $p^{2m}-1$ \\
$(p^{2m-1}+p^{m-1})(p-1)$  &  $p^{m-1}(p^{m}-1)(p^{m}-p+1)$ \\
$p^{2m-1}(p-1)-p^{m-1}$  &  $(p^{m}-1)(p-1)(p^{2m}-1)$ \\ \bottomrule
\end{tabular}
\end{table}
\noindent\textbf{Example 2.5.} Let $p=5$ and $m=1$, the code $C_{D_{2}}$ has parameters $[24,3,19]$ and weight enumerator $1+24x^{20}+96x^{19}+4x^{24}.$ This code is optimal.

\noindent\textbf{Example 2.6.} Let $p=3$ and $m=2$, the code $C_{D_{2}}$ has parameters $[80,6,51]$ and weight enumerator $1+480x^{51}+168x^{60}+80x^{54}.$ This code is optimal.

\section{Proofs of The Main Results}
Our task are to prove Theorem 2.1 and 2.3. Before doing this, we need to define a constant as follows. Let
\begin{equation}\label{4}
  n_{1}=|\{x\in F_{q}^{*}: Tr^{s}_{e}(x^{p^{m}+1})=0\}|,
\end{equation}
where $Tr^{s}_{e}(x)$ is the general trace function.
To prove Theorem 2.1 and 2.3, we also define the following parameter
\begin{equation*}
  N_{\beta}=|\{x\in F_{q}^{*}:Tr^{s}_{e}(x^{p^{m}+1})=0,~Tr^{s}_{1}(\beta x)=0\}|,
\end{equation*}
where $\beta \in F_{q}^{*}$. By definition and the basic facts of additive characters, for any $\beta \in F_{q}^{*}$, we have
  $$N_{\beta}=\frac{1}{p^{e+1}}\sum_{x\in F_{q}^{*}}(\sum_{\lambda\in F_{p^{e}}}\zeta_{p}^{Tr_{1}^{s}(\lambda x^{p^{m}+1})})(\sum_{y\in F_{p}}\zeta_{p}^{Tr^{s}_{1}(y\beta x)})
    $$
    $$=\frac{1}{p^{e+1}}(q+\sum_{x\in F_{q}}\sum_{\lambda\in F_{p^{e}}^{*}}\zeta_{p}^{Tr_{1}^{s}(\lambda x^{p^{m}+1})}+~~~~~~~~$$
\begin{equation}\label{4}
\sum_{\lambda\in F_{p^{e}}^{*}}\sum_{y\in F_{p}^{*}}\sum_{x\in F_{q}}\zeta_{p}^{Tr_{1}^{s}(\lambda x^{p^{m}+1})+Tr^{s}_{1}(y\beta x)})-1.
\end{equation}

Let $A=\sum\limits_{x\in F_{q}}(\sum\limits_{\lambda\in F_{p^{e}}^{*}}\zeta_{p}^{Tr_{1}^{s}(\lambda x^{p^{m}+1})})$ and
$B=\sum\limits_{\lambda\in F_{p^{e}}^{*}}\sum\limits_{y\in F_{p}^{*}}\sum\limits_{x\in F_{q}}\zeta_{p}^{Tr_{1}^{s}(\lambda x^{p^{m}+1})+Tr^{s}_{1}(y\beta x)}.$

Thus, we have the following lemmas.

\noindent\textbf{Lemma 3.1.} Let $s=2m$, $m=et$, $\lambda\in F_{p^{e}}^{*}$ and $\beta\in F_{q}$. Then
$$\sum_{x\in F_{q}}\zeta_{p}^{Tr^{m}_{1}(\lambda x^{p^{m}+1})+Tr^{s}_{1}(\beta x)}=-p^{m}\zeta_{p}^{Tr^{m}_{1}(-\lambda^{-1}\beta^{p^{m}+1})}.$$

\noindent\textbf{Proof.} By the basic facts of trace functions[18, Corollary 4], we have
$$ \sum_{x\in F_{q}}\zeta_{p}^{Tr^{m}_{1}(\lambda x^{p^{m}+1})+Tr^{s}_{1}(\beta x)} = \sum_{x\in F_{q}}\zeta_{p}^{Tr^{m}_{1}(\lambda x^{p^{m}+1}+\beta^{p^{m}} x^{p^{m}}+\beta x)}
$$ $$ = \sum_{x\in F_{q}}\zeta_{p}^{Tr^{m}_{1}(\lambda(x+\delta)^{p^{m}+1}-\lambda\delta^{p^{m}+1})}
=
\zeta_{p}^{Tr^{m}_{1}
(-\lambda\delta^{p^{m}+1})}((p^{m}+1)$$
  $\sum\limits_{z\in F_{p^{m}}^{*}}\zeta_{p}^{Tr^{m}_{1}(\lambda z)}+1)
   = -p^{m}\zeta_{p}^{Tr^{m}_{1}(-\lambda\delta^{p^{m}+1})},\\
$
where $\beta=\lambda\delta^{p^{m}}$(thus $\delta^{p^{m}+1}=\frac{\beta^{p^{m}+1}}{\lambda^{2}}$). So this completes the proof of this Lemma.\qed

\noindent\textbf{Lemma 3.2.} Let $s=2m$ and $e|m$. Then $A=\sum\limits_{\lambda\in F^{*}_{p^{e}}}\sum\limits_{x\in F_{q}}\zeta_{p}^{Tr^{s}_{1}(\lambda x^{p^{m}+1})}=-(p^{e}-1)p^{m}$ and the length $n$ of the $C_{D_{1}}$ is $p^{2m-e}-(p^{e}-1)p^{m-e}-1$.

\noindent\textbf{Proof.} According to Lemma 3.1, we could easily obtain the following result.
\begin{eqnarray*}
  A &=& \sum_{\lambda\in F^{*}_{p^{e}}}\sum_{x\in F_{q}}{Tr^{s}_{1}(\lambda x^{p^{m}+1})} \\
   &=& \sum_{\lambda\in F_{p^{e}}^{*}}(\sum_{z\in F^{*}_{p^{m}}}(p^{m}+1)\zeta_{p}^{Tr^{m}_{1}(\lambda z)}+1) \\
   &=& -(p^{e}-1)p^{m}.
\end{eqnarray*}
Combining (4) and the above result, we have the length $n=\frac{1}{p^{e}}(q+A)-1=p^{2m-e}-(p^{e}-1)p^{m-e}-1.$\qed

\noindent\textbf{Lemma 3.3.} Let $s=2m$, $m=et$, then

\begin{eqnarray*}
B=
\left\{ {{\begin{array}{ll}
 {-(p-1)(p^{e}-1)p^{m}}, & {\textrm{if}\mbox{ } Tr^{m}_{e}(\beta^{p^{m}+1})=0},\\
 {(p-1)p^{m}}, & {\textrm{if}\mbox{ } Tr^{m}_{e}(\beta^{p^{m}+1})\neq0}. \\
 \end{array} }} \right .
\end{eqnarray*}

\noindent\textbf{Proof.}
From the map $x\rightarrow \frac{y}{\lambda}x$ and $\lambda\rightarrow \frac{y^{2}}{\lambda}$, we have
\begin{eqnarray*}
  B &=& \sum_{y\in F_{p}^{*}}\sum_{\lambda\in F_{p^{e}}^{*}}\sum_{x\in F_{q}}\zeta_{p}^{Tr_{1}^{s}(\lambda (x^{p^{m}+1}+\beta x))}.
\end{eqnarray*}
By Lemma 3.1, we have
\begin{eqnarray*}
  B &=& -p^{m}(p-1)\sum_{\lambda\in F_{p^{e}}^{*}}\zeta_{p}^{Tr_{1}^{e}(\lambda Tr_{e}^{m}(\beta^{p^{m}+1}))} \\
   &=& \left\{ {{\begin{array}{ll}
 {-(p-1)(p^{e}-1)p^{m}}, & {\textrm{if}\mbox{ } Tr^{m}_{e}(\beta^{p^{m}+1})=0},\\
 {(p-1)p^{m}}, & {\textrm{if}\mbox{ } Tr^{m}_{e}(\beta^{p^{m}+1})\neq0}. \\
 \end{array} }} \right . \qed
\end{eqnarray*}

\noindent\textbf{The Proof of Theorem 2.1}

According to Lemma 3.2, the length of a codeword in $C_{D_{1}}$ is
$$n_{1}=p^{2m-e}-(p^{e}-1)p^{m-e}-1.$$
It follows from (5), Lemma 3.2 and Lemma 3.3 that we have
$$wt(c_{\beta})\in \{p^{2m-e}-p^{2m-e-1},(p^{2m-e-1}-p^{m-1})(p-1)\},$$
and the code $C_{D_{1}}$ has all the two weights in the set above.

Define
$w_{1}=p^{2m-e}-p^{2m-e-1}$, $w_{2}=(p^{2m-e-1}-p^{m-1})(p-1)$.
By Lemma 3.2, we have
\begin{eqnarray*}
  A_{w_{1}} &=& p^{2m-e}-(p^{e}-1)p^{m-e}-1, \\
  A_{w_{2}} &=& (p^{e}-1)(q+p^{m})/p^{e}.
\end{eqnarray*}\qed

\noindent\textbf{The Proof of Theorem 2.3}

Combining (3), Lemma 3.1 and Lemma 3.3, we obtain the following results.
\begin{eqnarray*}
  wt(c_{(\gamma,\beta)}) &=& q-p^{-1}\sum_{y\in F_{p}}\sum_{x\in F_{q}}\zeta_{p}^{y(Tr^{m}_{1}(\gamma x^{p^{m}+1})+Tr^{s}_{1}(\beta x))}
\end{eqnarray*}
\begin{eqnarray*}
=\left\{ {{\begin{array}{ll}
 {p^{2m-1}(p-1)}, & {\textrm{if}\mbox{ } \gamma=0,\beta\neq 0},\\
 {(p^{2m-1}+p^{m-1})(p-1)}, & {\textrm{if}\mbox{ } Tr^{m}_{1}(\gamma^{-1}\beta^{p^{m}+1})=0}. \\
 {p^{2m-1}(p-1)-p^{m-1}}, & {\textrm{if}\mbox{ } Tr^{m}_{1}(\gamma^{-1}\beta^{p^{m}+1})\neq0}.
 \end{array} }} \right .
\end{eqnarray*}

Let
$w_{1}=p^{2m-1}(p-1)$, $w_{2}=(p^{2m-1}+p^{m-1})(p-1)$, $w_{3}=p^{2m-1}(p-1)-p^{m-1}$. We determine the number $A_{w_{i}}$ of codewords with weight $w_{i}$ in $C_{D_{2}}$. It is possible to prove the minimum weight of the dual code $C^{\bot}_{D_{2}}$ is at least 3. Therefore, the first three Pless Power Moment lead to the following system of equations:
\begin{eqnarray*}
\left\{ {{\begin{array}{ll}
 {A_{w_{1}}+A_{w_{2}}+A_{w_{3}}=p^{3m}-1}, \\
 {w_{1}A_{w_{1}}+w_{2}A_{w_{2}}+w_{3}A_{w_{3}}=p^{3m-1}n_{2}(p-1)}, \\
 {w_{1}^{2}A_{w_{1}}+w_{2}^{2}A_{w_{2}}+w^{2}_{3}A_{w_{3}}=p^{3m-2}n_{2}(p-1)(n_{2}p-n_{2}+1)},
 \end{array} }} \right .
\end{eqnarray*}
where $n_{2}=p^{2m}-1$. Solving the system of equations yields the weight distribution in \noindent\textbf{Table III}.\qed

\section {The Duals of The Codes $C_{D_{1}}$ and $C_{D_{2}}$}

In this section, for the duals $C^{\perp}_{D_{1}}$ and $C^{\perp}_{D_{2}}$, we have the following two theorems.

\noindent\textbf{Theorem 4.1.}
Let $d^{\perp}_{1}$ denote the minimum distance of the $C^{\perp}_{D_{1}}$. The definition of $C_{D_{1}}$ can be found in Theorem 2.1. Then $2\leq d_{1}^{\perp}\leq 4$, $d_{1}^{\perp}=3$ if $p=2$ and $m\geq 3$.

\noindent\textbf{Proof.}
Clearly, $D_{1}$ does not contain the zero element of $F_{q}$, the minimum distance of $C^{\perp}_{D_{1}}$ cannot be one. Besides, $d_{1}^{\perp}$ is at most 4 due to the Sphere Packing Bound. Hence, we have $2\leq d_{1}^{\perp}\leq4$.

If $p=2$, the minimum distance $C^{\perp}_{D_{1}}$ cannot be 2, Since $D_{1}$ is not a multiset, any two elements $d_{i}$ and $d_{j}$ of $D_{1}$ must be distinct if $i\neq j$.

$D_{1}=\{x\in F_{q}^{*}:Tr^{s}_{e}(x^{p^{m}+1})=0\}$. Obviously, $F_{p^{e}}^{*}\subset D_{1}$. For any two distinct elements $a,b\in F_{p^{e}}^{*}\subset D_{1}$, we have $a+b\in F_{p^{e}}^{*}\subset D_{1}$. Besides, if $m\geq 3$, we have $2^{2m-e}+2^{m-e}-2^{m}\geq 2m-2$. Hence, the minimum distance of $C^{\perp}_{D_{1}}$ is 3.\qed

\noindent\textbf{Example 4.1.} Let $(m,e)=(2,1)$ and $p=3$. Then the code $C^{\perp}_{D_{1}}$ has parameters $[20,16,3]$ and is optimal.

\noindent\textbf{Example 4.2.} Let $p=2$, $m=3$ and $e=1$. Then the code $C^{\perp}_{D_{1}}$ has parameters $[27,21,3]$ and is almost optimal.  This code is optimal due to the Griesmer bound since the optimal linear code with length 27 and dimension 21 has minimum weight 4.

\noindent\textbf{Theorem 4.2.}
Let $d^{\perp}_{2}$ denote the minimum distance of the $C^{\perp}_{D_{2}}$. The code of $C_{D_{2}}$ is defined in equation (3). Then $3\leq d^{\perp}_{2}\leq 4$. Furthermore, in the special case of $p=3$, let $c=(c_{1},c_{2},\cdots,c_{n_{2}})$ be a codeword of $C^{\perp}_{D_{2}}$ with the minimum weight. Then $d^{\perp}_{2}=4$ if there exist three nonzero components $c_{i},c_{j},c_{k}$ of $c$ such that $c_{i}=c_{j}=c_{k}=1$ or 2, for some positive integers $i,j,k\in \{0,1,\cdots,n_{2}-1\}$. Otherwise, $d_{2}^{\perp}=3$.

\noindent\textbf{Proof.}
Clearly, $d_{2}^{\perp}\geq 2$. Now we could prove that $d_{2}^{\perp}\neq2$. By the definition of $C_{D_{2}}$, $d^{\perp}_{2}=2$ if and only if there are two distinct elements $x_{1},x_{2}\in F_{p^{2m}}^{*}$ and $c_{1},c_{2}\in F_{p}^{*}$ such that
$$c_{1}(Tr^{m}_{1}(\gamma x_{1}^{p^{m}+1})+Tr^{s}_{1}(\beta x_{1}))+c_{2}(Tr^{m}_{1}(\gamma x_{2}^{p^{m}+1})+Tr^{s}_{1}(\beta x_{2}))$$
$$=Tr^{m}_{1}(\gamma (c_{1} x_{1}^{p^{m}+1}+c_{2} x_{2}^{p^{m}+1})+\beta^{p^{m}}(c_{1} x_{1}^{p^{m}}+c_{2}x^{p^{m}}_{2})+$$ $\beta( c_{1}x_{1}+c_{2}x_{2})) =0.\\$
for all $\gamma\in F_{p^{m}}$ and $\beta\in F_{q}$. This is equivalent to
\begin{eqnarray}\label{6}
 \left\{ {{\begin{array}{ll}
 {c_{1} x_{1}^{p^{m}+1}+c_{2} x_{2}^{p^{m}+1}=0}, \\
 {c_{1} x_{1}^{p^{m}}+c_{2} x^{p^{m}}_{2} = 0}, \\
 {c_{1} x_{1}+c_{2} x_{2}=0}. \\
\end{array} }} \right .
\end{eqnarray}

By the equations of (6), we have
$$c_{1}\frac{c_{2}^{2}}{c_{1}^{2}}x_{2}^{p^{m}+1}+c_{2}x_{2}^{p^{m}+1}
=\frac{(c_{2}^{2}+c_{1}c_{2})x_{2}^{p^{m}+1}}{c_{1}}=0.$$

Therefore, we have $c_{2}=0$ or $c_{2}=-c_{1}$, which is a contradiction to the facts that $c_{2}\in F_{p}^{*}$ and $x_{1}\neq x_{2}$, respectively.

As the minimum weight of any linear code with length $p^{2m}-1$ and dimension $3m$ is at most 4 due to the Sphere Packing Bound, we have $d_{2}^{\perp}\leq 4$. This completes the proof of the conclusion in the first part of this theorem.

Now we consider the special case that $p=3$. Obviously, $C^{\perp}_{D_{2}}$ has a codeword of weight three if and only if there are three pairwise distinct elements $x_{1},x_{2},x_{3}\in F_{3^{2m}}^{*}$ and three elements $c_{1},c_{2},c_{3}\in F_{3}^{*}$ such that
$$Tr^{m}_{1}(\gamma (c_{1}x_{1}^{3^{m}+1}+c_{2}x_{2}^{3^{m}+1}+c_{3}x_{3}^{3^{m}+1})+\beta^{3^{m}} (c_{1}x^{3^{m}}_{1}+
$$ $$c_{2} x^{3^{m}}_{2}+c_{3}x^{3^{m}}_{3})+\beta (c_{1} x_{1}+c_{2} x_{2}
  +c_{3}x_{3}))=0,$$
for all $\gamma\in F_{3^{m}}$ and $\beta\in F_{3^{2m}}$. This is equivalent to
\begin{eqnarray}\label{7}
 \left\{ {{\begin{array}{ll}
 {c_{1} x_{1}^{3^{m}+1}+c_{2} x_{2}^{3^{m}+1}+c_{3} x_{3}^{3^{m}+1}=0}, \\
 {c_{1} x_{1}^{3^{m}}+c_{2} x^{3^{m}}_{2}+c_{3} x^{3^{m}}_{3} = 0}, \\
 {c_{1} x_{1}+c_{2} x_{2}+c_{3} x_{3}=0}. \\
\end{array} }} \right .
\end{eqnarray}
 Without loss of generality, we only need to consider the following two subcases, since other situations are equivalent to the two subcases.

 1. We assume that $c_{1}=c_{2}=c_{3}=1$ or $c_{1}=c_{2}=c_{3}=2$, which is the first case.
 It then follows from the last equations of (7) that $x_{1}=-(x_{2}+x_{3}).$ We have
$$(-x_{2}-x_{3})^{3^{m}+1}+x_{2}^{3^{m}+1}+x_{3}^{3^{m}+1} = 2x_{2}^{3^{m}+1}+2x_{3}^{3^{m}+1}$$ $$ +x_{2}x_{3}^{3^{m}}+x_{3}x_{2}^{3^{m}}
  =2x_{2}^{3^{m}+1}+2x_{3}^{3^{m}+1}-2x_{2}x_{3}^{3^{m}}-2x_{3}x_{2}^{3^{m}}$$ $ = 2(x_{2}-x_{3})^{3^{m}+1}=0,
$\\
which is a contradiction. When $c_{1}=c_{2}=c_{3}=2$, the proof of this case is similar to $c_{1}=c_{2}=c_{3}=1.$

2. $c_{1}=c_{2}=1,c_{3}=-1$, other cases are similar to it.
From the last equations (7), we have $x_{3}=x_{1}+x_{2}$ and
  $$-(x_{1}+x_{2})^{3^{m}+1}+x_{1}^{3^{m}+1}+x_{2}^{3^{m}+1} = -x_{1}x_{2}^{3^{m}}-x_{1}x_{2}^{3^{m}}=$$
$x_{1}x_{2}(x_{1}^{3^{m}-1}+x_{2}^{3^{m}-1})=0.$\\
Clearly, it is possible that $(\frac{x_{2}}{x_{1}})^{3^{m}-1}=-1.$ Therefore, the proof of this theorem is now completed.\qed

\noindent\textbf{Example 4.3.}
Let $p=5$ and $m=1$. Then the code $C^{\perp}_{D_{2}}$ has parameters $[24,21,3]$ and is optimal.

\noindent\textbf{Example 4.4.}
Let $p=3$ and $m=2$, the code $C^{\perp}_{D_{2}}$ has parameters $[80,74,3]$ and is optimal.

\section {Conclusion}
In this paper, we generalized the construction of linear codes by Ding et al[3]. The general construction method can get linear codes with flexible lengths and dimensions. Besides,
linear codes over $F_{p}$ have wide applications which are used for the construction of secret sharing schemes[3] and authentication codes[15]. Let $w_{min}$ and $w_{max}$ denote the minimum and maximum nonzero Hamming weights of the code $C$. In order to obtain secret sharing with interesting access structures, we would like to have linear codes $C$ such that $\frac{w_{min}}{w_{max}}>\frac{p-1}{p}$[16].

Then for the code $C_{D_{1}}$ and $C_{D_{2}}$ of Theorem 2.1 and 2.3 we have
$$\frac{w_{min}}{w_{max}}=\frac{(p^{2m-e-1}-p^{m-1})(p-1)}{p^{2m-e}-p^{2m-e-1}}>\frac{p-1}{p}.$$
$$\frac{w_{min}}{w_{max}}=\frac{p^{2m-1}(p-1)-p^{m-1}}{(p^{2m-1}+p^{m-1})(p-1)}>\frac{p-1}{p}.$$

Hence, the linear codes $C_{D_{1}}$ and $C_{D_{2}}$ of this paper satisfy the condition that $\frac{w_{min}}{w_{max}}>\frac{p-1}{p}$ and can be employed to obtain secret sharing schemes with interesting access structures using the framework in [16].

\ifCLASSOPTIONcaptionsoff
  \newpage
\fi


\begin{thebibliography}{1}

\bibitem{pa} C. Carlet, C. Ding and J. Yuan, ``Linear codes from perfect nonlinear mappings and their secret sharing schemes," IEEE Trans. Inf. Theory, vol. 51, no. 6, pp. 2089-2102, Jun. 2005.

\bibitem{pa} J. Yuan, C. Carlet and C. Ding, ``The weight distribution of a class of linear codes from perfect nonlinear functions," IEEE Trans. Inf. Theory, vol. 52, no. 2, pp. 712-716, Feb. 2006.

\bibitem{pa} K. Ding and C. Ding, ``Binary linear codes with three weights," IEEE Trans. Inf. Theory, vol. 18, no. 11, pp. 1879-1882, Nov. 2014.

\bibitem{pa} K. Ding and C. Ding, ``A class of two-weight and three-weight codes and their applications in secret sharing," IEEE Trans. Inf. Theory, vol. 61, no. 11, pp. 5835-5842, Nov. 2015.

\bibitem{pa} Q. Wang, K. Ding and R. Xue, ``Binary linear codes with two weights," IEEE Trans. Inf. Theory, vol. 19, no. 7, pp. 1097-1100, Jul. 2015.
\bibitem{pa} Y. Qi, C. Tang and D. Huang, ``Binary linear codes with few weights," IEEE Commun. Lett., vol. 20, no. 2, pp. 208-211, Feb. 2016.

\bibitem{pa} C. Ding, J. Luo, and H. Niederreiter, ``Two weight codes punctured from irreducible cyclic codes," in Proc. 1st Int. Workshop coding Theory Cryptogr., 2008, pp. 119-124.

\bibitem{pa} K. Feng and J. Luo, ``Weight distribution of some reducible cyclic codes," Finite Fields Appl., 14 (2008) 390-409.


\bibitem{pa} Z. Heng and Q. Yue, ``A class of binary linear codes with at most three weights," IEEE Commun. Lett., vol. 19, no. 9, pp. 1488-1491, Sep. 2015.

\bibitem{pa21} C.Tang, N. Li, Y. Qi, Z. Zhou and T. Helleseth, ``Two-weight and three-weight linear codes from weakly regular bent function," arXiv:1507.0148v3.

\bibitem{pa} W. C. Huffman and V. Pless, ``Fundamentals of Error-Correcting Codes," Cambridge, U.K.: Cambridge Univ. Press, 2003.

\bibitem{pa} K. Feng and J. Luo, ``Weight distribution of some reducible cyclic codes," Finite Fields Appl., 14 (2008) 390-409.

\bibitem{pa} X. Zeng, L. Hu, W. Jiang, Q, Yue and X. Cao, ``The weight distribution of a class of p-ary cyclic codes," Finite Fields Appl., 16 (2010) 56-73.


\bibitem{pa} C. Ding, J. Yang, ``Hamming weight in irreducible cyclic codes," Discrete Math., 313 (4) (2013) 434-446.

\bibitem{pa4} C. Ding and X. Wang, ``A coding theory construction of new systematic authentication codes," Theoretical Comput. Sci., vol. 330, no. 1, pp. 81-99, Jan. 2005.

\bibitem{pa4} J. Yuan and C. Ding, ``Secret sharing schemes from two-weight codes," in Pro. R.C. Bose Centenary Symp., Discr. Math. Appl., KOlkata, India, Dec. 2002, p. 232.

\bibitem{pa} C. Li and Q. Yue, ``Weight distributions of cyclic codes with respect to pairwise coprime order elements," Finite Fields Appl., vol.28, pp. 94-114, Jul. 2014.

\bibitem{pa} T. Helleseth and A. Kholosha, ``Monomial and quadratic Bent functions over the finite fields of odd characteristic," IEEE Trans. Inf. Theory. vol. 52, no. 5, pp. 2018-2032, May. 2006.

\end{thebibliography}
\end{document}